\documentclass[a4paper,11pt]{scrartcl}

\usepackage[top=3cm,left=3cm,right=3cm,bottom=3cm]{geometry}
\usepackage[intlimits]{amsmath}
\usepackage{amsfonts,amssymb,amsthm}
\usepackage{hyperref}
\usepackage{enumerate}
\usepackage{braket}
\usepackage{graphicx}
\usepackage{caption}
\usepackage{subcaption}
\usepackage{color}
\usepackage{url}
\allowdisplaybreaks

\hypersetup{pdfstartview=FitH}

\newtheorem{theorem}{Theorem}[section]

\begin{document}

\title{The role of tessellation intersection in staggered quantum walks}
\author{Raqueline A. M. Santos}
\date{\small{Center for Quantum Computer Science, Faculty of Computing, University of Latvia} \\ 
\small{Raina bulv. 19, Riga, LV-1586, Latvia}\\
\small{\texttt{rsantos@lu.lv}}}
\maketitle

\begin{abstract}
The staggered quantum walk (SQW) model is defined by partitioning the graph into cliques, which are called polygons. 
We analyze the role that the size of the polygon intersection plays on the dynamics of SQWs on graphs.
We introduce two processes (intersection reduction and intersection expansion), that change the number of vertices in some intersection of polygons, and we compare the behavior of the SQW on the reduced or expanded graph in relation to the SQW on the original graph. 
We describe how the eigenvectors and eigenvalues of the evolution operators relate to each other. 
This processes can help to establish the equivalence between SQWs on different graphs and to simplify the analysis of SQWs.
We also show an example of a SQW on a graph that is not included in Szegedy's model, but which is equivalent to an instance of Szegedy's model after applying the intersection reduction. 
 
\end{abstract}

\section{Introduction}

The staggered quantum walk (SQW) model~\cite{Portugal:2015} has been actively studied in the last years and its relation with other quantum walk models has already been established. 
Ref.~\cite{Portugal:2015} showed that Szegedy's quantum walk model~\cite{Szegedy:2004} is included in the SQW model. Ref.~\cite{Portugal:2016} showed that many instances of coined QWs~\cite{Aharonov:1993, Aharonov:2001} can be cast into Szegedy's model and therefore into the SQW model, including the abstract search algorithm scheme~\cite{Ambainis:2005}. Ref.~\cite{Coutinho:2018} showed that the SQW model is able to provide a natural discretization of a continuous time quantum walk~\cite{Farhi:1998} for some special class of graphs.

In terms of physical implementation, Ref.~\cite{Portugal:2017} presented an extension of the SQW model that can be used for physical implementations in terms of time independent Hamiltonians. And Ref.~\cite{Jalil:2017} proposed an implementation with superconducting microwave resonators. In terms of algorithmic applications, spatial quantum search was analyzed on the two dimensional lattice~\cite{Portugal:2017a} and on hexagonal lattices~\cite{Chagas:2018}. The quantum algorithm for element distinctness was analyzed using the SQW model by~\cite{Portugal:2018} obtaining optimal values for some critical parameters of Ambainis' quantum algorithm~\cite{Ambainis:2004}

\begin{figure}[!htb]
  \centering
\includegraphics[scale=1.5]{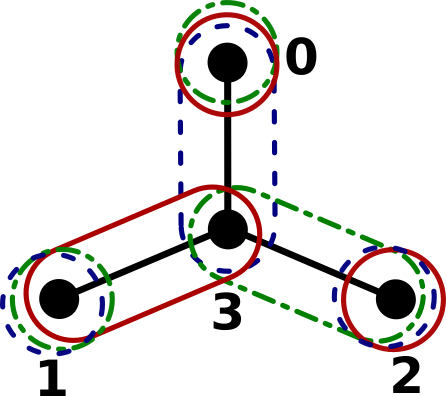} 
\caption{A 3-tessellation cover for the star graph $S_3$. The blue tessellation contains polygons $\{0,3\},\{1\}$ and $\{2\}$. The red tessellation contains polygons $\{0\},\{1,3\}$ and $\{2\}$. The green tessellation contains polygons $\{0\},\{1\}$ and $\{2,3\}$. Each polygon is a clique. The union of all polygons covers all the edges of the graph.}
\label{fig:star}
\end{figure}

A SQW on a graph is defined by a graph tessellation cover. A {\em tessellation} is a partitioning of the graph into cliques, called {\em polygons}. A {\em tessellation cover} is a set of tessellations that cover all the edges of the graph. We say that an edge belongs to a tessellation if both  of its endpoints belong to the same polygon. See Fig.~\ref{fig:star} for an example of a tessellation cover of a graph. Ref.~\cite{Abreu:2018} showed that the
problem of deciding whether a graph is $t$-tessellable is NP-complete.

The first step to define a SQW on a graph is to find a tessellation cover for it. 
With the tessellations in hands, we associate a unit vector to each polygon. Then, a local operator is defined for each tessellation as a reflection through the space spanned by the polygons. And the evolution operator is the product of the local operators for each tessellation. Let us see an example. 
Suppose we have the star graph $S_3$ in Fig.~\ref{fig:star} covered by 3 tessellations: 
$\mathcal{T}_{\textrm{blue}} = \{\{0,3\},\{1\},\{2\}\}$, $\mathcal{T}_{\textrm{red}} = \{\{0\},\{1,3\},\{2\}\}$ and $\mathcal{T}_{\textrm{green}} =\{\{0\},\{1\},\{2,3\}\}$.
The Hilbert space associated to this graph has dimension 4. We can associate unit vectors to the polygons in each tessellation.  For the blue tessellation, define
\begin{align}\label{eq:sqwstarblue}
&\ket{\alpha_0} = \frac{1}{\sqrt{2}}\left(\ket{0}+i\ket{3}\right),\quad \ket{\alpha_1} = \ket{1},\quad \ket{\alpha_2} = \ket{2};
\end{align}
for the red tessellation,
\begin{align}\label{eq:sqwstarred}
&\ket{\beta_0} = \ket{0},\quad \ket{\beta_1} = \frac{1}{\sqrt{2}}\left(\ket{1}+\ket{3}\right),\quad \ket{\beta_2} = \ket{2};
\end{align}
and for the green tessellation,
\begin{align}\label{eq:sqwstargreen}
&\ket{\gamma_0} = \ket{0},\quad \ket{\gamma_1} = \ket{1},\quad \ket{\gamma_2} = \frac{1}{\sqrt{2}}\left(\ket{2}+\ket{3}\right).
\end{align}
The amplitudes in those states can be arbitrarily defined, as long as the states remain unitary. From these states, we can generate local operators, which allows a particle to move only to neighboring vertices inside its polygon,
$$
U_{\textrm{blue}} = 2\sum_{j = 0}^2\ket{\alpha_j}\bra{\alpha_j}-I,\quad U_{\textrm{red}} = 2\sum_{j = 0}^2\ket{\beta_j}\bra{\beta_j}-I,\quad 
U_{\textrm{green}} = 2\sum_{j = 0}^2\ket{\gamma_j}\bra{\gamma_j}-I.
$$
Finally, the evolution operator is given by the product of the local operators, 
\begin{align}\label{eq:start3U}
U = U_{\textrm{green}} U_{\textrm{red}}U_{\textrm{blue}}.
\end{align}
And now we have completely defined a SQW on $S_3$.

In this paper, we are interested on what happens when we have more than one vertex in the intersection of polygons (the SQW in Fig.~\ref{fig:star}, for example, has one vertex in all intersections). We introduce two processes ({\em intersection expansion} and {\em intersection reduction}) that change the number of vertices in the intersection. We analyze the dynamics of the SQWs on the original graph and on the reduced or expanded graph. We show that the SQW on the graph obtained by these processes are equivalent to the SQW on the original graph, if some assumptions are made for the vertices in the intersection. 

Additionally, we analyze what happens to the eigenvalues and eigenvectors of the evolution operators. Both the original and reduced or expanded SQW will share some eigenvalues and their eigenvectors can be expressed in terms of each other. The SQW with bigger space have additional eigenvectors which can be described explicitly. These eigenvectors are associated with eigenvalue $+1$ if the number of tessellations is even. Otherwise, they are associated with eigenvalue $-1$.
This study can help to better understand this model and to find equivalence between SQWs on different graphs. Moreover, it can simplify the analysis of SQWs which have more than one vertex in some polygons intersection.

Our paper is divided as follows. The intersection expansion process is explained in Sec.~\ref{sec:exp}, followed by the intersection reduction in Sec.~\ref{sec:red}. Examples are drawn in both sections.
In Sec.~\ref{sec:results}, we summarize and combine the results obtained in the previous sections. An example of the spatial search on the two-dimensional lattice is presented. The conclusions are presented in Sec.~\ref{sec:conc}.

\section{Intersection Expansion}\label{sec:exp}

\begin{figure}[!htb]
  \centering

  \begin{subfigure}[t]{0.45\textwidth}
  \centering
\includegraphics[scale=1.5]{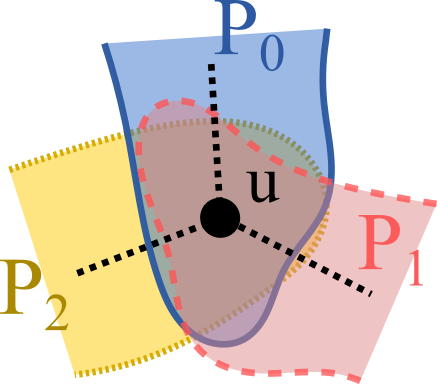}  
     \caption{One vertex is in the intersection of polygons $P_0, P_1, P_2$.}
     \label{fig:graphu}
 \end{subfigure}
 \begin{subfigure}[t]{0.45\textwidth}
 \centering
 \includegraphics[scale=1.5]{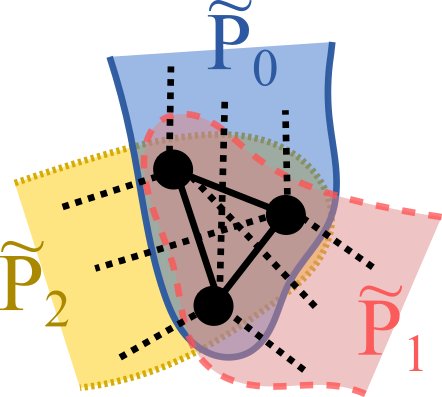}  
     \caption{Three vertices are in the intersection of polygons $\tilde{P}_0, \tilde{P}_1, \tilde{P}_2$.}
     \label{fig:graphuexpanded}
 \end{subfigure}
  
  \caption{Example of SQWs with $3$ tessellations and different number of vertices in some polygons intersection.}
  \label{fig:reduction2}
\end{figure}

Suppose we have a SQW on a graph $G=(V,E)$ with $l$ tessellations. Moreover, suppose $G$ has a vertex $u$ which is the only vertex in the intersection of some polygons. See an example with 3 tessellations in Fig.~\ref{fig:graphu}.
Let $P_j$ ($j=0,\dots ,l-1$) be the polygons which contain vertex $u$.
The state $\ket{P_j} \in \mathcal{H}^{|V|}$ induced by polygon $P_j$ is
\begin{equation}
\ket{P_j} = \sum_{v\in P_j}c_v^{(j)}\ket{v}, {\textrm{where }}\sum_{v\in P_j}|c_v^{(j)}|^2=1.
\end{equation}
The reflection operator associated to each tessellation is
\begin{equation}
U_{j} = 2\left(\ket{P_j}\bra{P_j}+R\right) -I.
\end{equation} 
where $R$ contains the sum of the outer products of the states associated to the remaining polygons.
The evolution operator $U$ is a product of all reflection operators generated by the polygons in each tessellation, $U = U_1U_2 \dots U_l$.

We ``expand'' vertex $u$ into a $k$-clique, such that the intersection of polygons has more than one vertex. That is, we transform a graph in Fig.~\ref{fig:graphu} into a graph in Fig.~\ref{fig:graphuexpanded}, for example. Let us label the vertices of the $k$-clique as $\{0,1,\dots,k-1\}$.
Call the new graph $\tilde{G} = (\tilde{V},\tilde{E})$, where $\tilde{V}=  \{0,1,\dots,k-1\}\cup V\backslash\{u\}$.
Each vertex in the clique is connected with each neighbour of the original vertex. That is, for any $v\in V$ if $v\sim u$, then $v\sim \{0,\dots,k-1\}$. 
The $k$-clique will belong to the same polygons as the original vertex $u$. Since we are inserting cliques and maintaining all the connections from the previous graph, the tessellations will remain valid. Each polygon will still contain a clique. 

Let us associate a unitary state, $\ket{\tilde{u}} \in \mathcal{H}^{|\tilde{V}|}$, to the $k$-clique,
\begin{equation}
\ket{\tilde{u}} = \sum_{j=0}^{k-1}u_j\ket{j}, {\textrm{where }} \sum_{j=0}^{k-1}|u_j|^2 = 1.
\end{equation}
This state will determine how we describe the amplitudes of the $k$-clique in each polygon. Define the state induced by polygon $\tilde{P}_ {j}$  as
\begin{equation}
\ket{\tilde{P}_j} = \sum_{v\in V\backslash\{u\}}c_v^{(j)}\ket{v}+c_{u}^{(j)}\ket{\tilde{u}}.
\end{equation} 
Notice that $\ket{\tilde{P}_j} \in \mathcal{H}^{|\tilde{V}|}$ is unitary. In the same way as before, we can generate the reflection operators, $\tilde{U}_{j}$, for each tessellation, and the evolution operator $\tilde{U} = \tilde{U}_1\tilde{U}_2\dots \tilde{U}_l$.

Let us see how is the action of $U_{j}$ on a generic state, 
\begin{equation}
\ket{\psi}=\sum_{v\in V}a_v\ket{v}.
\end{equation}
We have,
\begin{equation}\label{eq:Uipsi}
U_j\ket{\psi} = \sum_{v\in P_j}(2mc_v^{(j)}-a_v)\ket{v} + U_j\left(\sum_{v\notin P_j}a_v\ket{v}\right),{\textrm{where }} m=\sum_{v\in P_j}\overline{c_v^{(j)}}a_v,
\end{equation}
and $\overline{c_v^{(j)}}$ means the complex conjugate of $c_v^{(j)}$. 
Remember that $U_j$ acts locally on each polygon.

Now, define
\begin{equation}\label{eq:tildepsi}
\ket{\tilde{\psi}} = \sum_{v\in V\backslash\{u\}}a_v\ket{v}+a_{u}\ket{\tilde{u}}.
\end{equation}
Then,
\begin{equation}
\tilde{U}_{j}\ket{\tilde{\psi}} = \sum_{v\in P_j\backslash\{u\}}(2\tilde{m}{c_v^{(j)}}-{a_v})\ket{v}+\sum_{v=0}^{k-1}(2\tilde{m}c_u^{(j)}-a_u)u_v\ket{v}+\tilde{U}_j\left(\sum_{v\notin \tilde{P}_j}a_v\ket{v}\right),
\end{equation}
where
\begin{equation}
\tilde{m}=\sum_{v\in P\backslash\{u\}}\overline{c_v^{(j)}}{a_v}+\sum_{v=0}^{k-1}\overline{c_u^{(j)}u_v}a_uu_v =\sum_{v\in P\backslash\{u\}}\overline{c_v^{(j)}}a_v+\overline{c_u^{(j)}}u_u\sum_{v=0}^{k-1}|u_v|^2 = m.
\end{equation}
Therefore,
\begin{equation}\label{eq:tildeUipsi}
\tilde{U}_{j}\ket{\tilde{\psi}} = \sum_{v\in P_i\backslash\{u\}}(2m{c_v^{(j)}}-{a_v})\ket{v}+(2mc_u^{(j)}-a_u)\ket{\tilde{u}}+\tilde{U}_j\left(\sum_{v\notin {P}_j}a_v\ket{v}\right).
\end{equation}

We are interested in the part where the expansion was made. The rest of the graph remains the same, in both cases, and the application of the reflection operator in the other polygons which does not contain the vertex $u$ or the clique will remain the same. This means that the last term of Eq.~(\ref{eq:Uipsi}) will be the same as the last term of Eq.~(\ref{eq:tildeUipsi}), despite of a dimensionality difference.

From Eqs. (\ref{eq:Uipsi}) and (\ref{eq:tildeUipsi}), we can see that the amplitudes of the vertices belonging to $P_i\backslash\{u\}$ are the same in both equations. For the vertices in the intersection of polygons, we have $\braket{\tilde{u}|\tilde{U}_j|\tilde{\psi}} = \braket{{u}|{U}_j|{\psi}}$. The amplitude of $u$ is divided  into the vertices of the clique depending on the amplitudes of $\ket{\tilde{u}}$.
This implies that when we apply the evolution operator $U$/$\tilde{U}$ to some state $\ket{\psi}$/$\ket{\tilde{\psi}}$, the amplitudes of the vertices belonging to  $P_i\backslash\{u\}$ will be the same and the probability of obtaining vertex $u$ after measurement (in the computational basis) is the same as the probability of obtaining one of the vertices in the clique. It is important to notice that this is valid only if $\ket{\tilde{\psi}}$ has amplitudes multiples of $\ket{\tilde{u}}$ for the vertices in the intersection, as defined in Eq.~(\ref{eq:tildepsi}). More formally, we can write
\begin{equation}\label{eq:Upsi}
U\ket{\psi} = \sum_{v\in V}\gamma_v\ket{v},
\end{equation}
then 
\begin{equation}\label{eq:tildeUpsi}
\tilde{U}\ket{\tilde{\psi}} = \sum_{v\in V\backslash\{u\}}\gamma_v\ket{v}+\gamma_u\ket{\tilde{u}}.
\end{equation}

\subsection{Eigenvectors and eigenvalues of the evolution operators}\label{sec:spectra}
It is easy to identify part of the eigenvectors and eigenvalues of $\tilde{U}$.
Let $\ket{\lambda} \in \mathcal{H}^{|V|}$,
\begin{equation}\label{eq:lambda}
\ket{\lambda} = \sum_{v\in V}\beta_v\ket{v},
\end{equation}
 be a $\lambda$-eigenvector of $U$, that is, $U\ket{\lambda}=\lambda\ket{\lambda}$. Define $\ket{\tilde{\lambda}}\in \mathcal{H}^{|\tilde{V}|}$ such that
\begin{equation}\label{eq:tildelambda}
\ket{\tilde{\lambda}} = \sum_{v\in V\backslash\{u\}}\beta_v\ket{v}+\beta_u\ket{\tilde{u}}.
\end{equation}
Then, from Eqs.~(\ref{eq:Upsi}) and (\ref{eq:tildeUpsi}), we have
\begin{equation}
\tilde{U}\ket{\tilde{\lambda}} = \tilde{U}\left(\sum_{v\in V\backslash\{u\}}\beta_v\ket{v}+\beta_u\ket{\tilde{u}} \right) = \sum_{v\in V\backslash\{u\}}\lambda\beta_v\ket{v}+\lambda\beta_u\ket{\tilde{u}} = \lambda\ket{\tilde{\lambda}}.
\end{equation}
That is, $\ket{\tilde{\lambda}}$ is a $\lambda$-eigenvector of $\tilde{U}$. All eigenvalues of $U$ are eigenvalues of $\tilde{U}$. However, $\tilde{U}$ has bigger dimension and it has additional $k-1$ eigenvectors. The \emph{ansatz} is that a state $\ket{\phi}$ such that $\braket{\phi|v} = 0$ if $v \in V\backslash\{u\}$ and $\braket{\tilde{P}_j|\phi} = 0$ for all $j$, will be an eigenvector of $\tilde{U}$. Notice that  $\tilde{U}_{j}\ket{\phi}$ will simply flip the sign of the amplitudes of $\ket{\phi}$. Therefore, $\ket{\phi}$ can be a $+1$-eigenvector of $\tilde{U}$ if the number or tessellations is even, because $\tilde{U}$ will flip the sign of the amplitudes an even number of times. Or $\ket{\phi}$ can be a $-1$-eigenvector if the number of tessellations is odd.
Having more than one vertex in the intersection allows us to easily find a set of orthonormal eigenvectors of this kind. Let $\ket{\phi} = \sum_{v=0}^{k-1}b_v\ket{v}$. Then
\begin{equation}\label{eq:braketPphi}
\braket{\tilde{P}_j|\phi} = 0 \Rightarrow \sum_{v=0}^{k-1}\overline{\tilde{c}_v^{(j)}}b_v = 0 \Rightarrow \sum_{v=0}^{k-1}\overline{c_u^{(j)}u_v}b_v = 0 \Rightarrow \sum_{v=0}^{k-1}\overline{u_v}b_v = 0.
\end{equation}
We can find a set of $k-1$ orthonormalized vectors that satisfy Eq.~(\ref{eq:braketPphi}).
Define
\begin{equation}\label{eq:nuj}
\ket{\nu_j} = \frac{1}{\gamma_j}\left(\frac{\overline{u_j}}{\gamma_{j-1}}\sum_{v=0}^{j-1}u_v\ket{v}-\gamma_{j-1}\ket{j} \right),
\end{equation}
where $1\leq j \leq k-1$ and $\gamma_j = \sqrt{\sum_{v=0}^j|u_v|^2}$. It is easy to check that $\ket{\nu_j}$ satisfy Eq.~(\ref{eq:braketPphi}) and $\braket{\nu_i|\nu_j} = 0$, for $i\neq j$. Moreover, $\braket{\nu_j|\tilde{u}} = 0$, for all $j$, which means that the eigenvectors $\ket{\nu_j}$ are orthogonal to the other eigenvectors of $\tilde{U}$ described by Eq.~(\ref{eq:tildelambda}).

\subsection{Example - Expansion on the star graph $S_3$}\label{sec:expexample}
\begin{figure}[!htb]
  \centering
\includegraphics[scale=1.5]{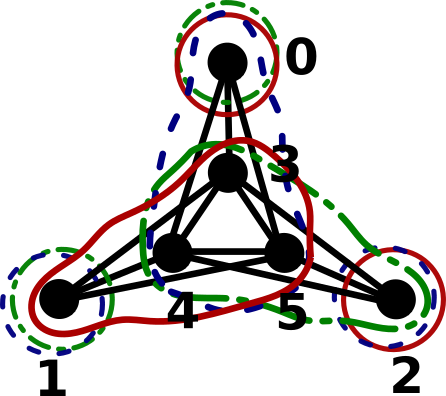} 
\caption{A 3-tessellation cover for the star graph $S_3$.}
\label{fig:star2}
\end{figure}

Let us see a small example. Consider we expanded vertex $3$ from the star graph $S_3$ in Fig.~\ref{fig:star} by a 3-clique. The result is shown in Fig.~\ref{fig:star2}.
Let us associate a unit vector to the 3-clique,
$$
\ket{\tilde{u}} = u_3\ket{3}+u_4\ket{4}+u_5\ket{5}.
$$
Let us choose the uniform superposition by making $u_3= u_4 = u_5 = \frac{1}{\sqrt{3}}$.
Then, following the same structure as in Eqs.~(\ref{eq:sqwstarblue})-(\ref{eq:sqwstargreen}), the unit vectors associated to the polygons in each tessellation are
$$
\ket{\tilde{\alpha}_0} = \frac{1}{\sqrt{2}}\left(\ket{0}+i\ket{\tilde{u}}\right), \ket{\tilde{\alpha}_1} = \ket{1}, \ket{\tilde{\alpha}_2} = \ket{2},
$$
$$
\ket{\tilde{\beta}_0} = \ket{0}, \ket{\tilde{\beta}_1} = \frac{1}{\sqrt{2}}\left(\ket{1}+\ket{\tilde{u}}\right), \ket{\tilde{\beta}_2} = \ket{2},
$$
$$
\ket{\tilde{\gamma}_0} = \ket{0}, \ket{\tilde{\gamma}_1} = \ket{1}, \ket{\tilde{\gamma}_2} = \frac{1}{\sqrt{2}}\left(\ket{2}+\ket{\tilde{u}}\right).
$$
We simply substituted $\ket{3}$ in Eqs.~(\ref{eq:sqwstarblue})-(\ref{eq:sqwstargreen}) by $\ket{\tilde{u}}$.
The local operators are
$$
\tilde{U}_{\textrm{blue}} = 2\sum_{j = 0}^2\ket{\tilde{\alpha}_j}\bra{\tilde{\alpha}_j}-I, \tilde{U}_{\textrm{red}} = 2\sum_{j = 0}^2\ket{\tilde{\beta}_j}\bra{\tilde{\beta}_j}-I, 
\tilde{U}_{\textrm{green}} = 2\sum_{j = 0}^2\ket{\tilde{\gamma}_j}\bra{\tilde{\gamma}_j}-I
$$
and the evolution operator $\tilde{U} = \tilde{U}_{\textrm{green}} \tilde{U}_{\textrm{red}}\tilde{U}_{\textrm{blue}}$.
Finding the spectral decompostion of $\tilde{U}$, we obtain
\begin{align*}
\tilde{U} = i\ket{\tilde{\lambda}^{+i}}\bra{\tilde{\lambda}^{+i}}-i\ket{\tilde{\lambda}^{-i}}\bra{\tilde{\lambda}^{-i}}+\ket{\tilde{\lambda}^{+1}}\bra{\tilde{\lambda}^{+1}}-\left(\ket{\tilde{\lambda}^{-1}}\bra{\tilde{\lambda}^{-1}}+\ket{\nu_1}\bra{\nu_1}+\ket{\nu_2}\bra{\nu_2}\right),
\end{align*}
where
\begin{align*}
&\ket{\tilde{\lambda}^{+i}} = \frac{1}{2}\begin{bmatrix}\begin{array}{r}
-1\\-1\\i\\\frac{1}{\sqrt{3}}\\\frac{1}{\sqrt{3}}\\\frac{1}{\sqrt{3}}\end{array}
\end{bmatrix},\quad
\ket{\tilde{\lambda}^{-i}} = \frac{1}{2}\begin{bmatrix}\begin{array}{r}
1\\-1\\-i\\\frac{1}{\sqrt{3}}\\\frac{1}{\sqrt{3}}\\\frac{1}{\sqrt{3}}\end{array}
\end{bmatrix},\quad
\ket{\tilde{\lambda}^{+1}} = \frac{1}{2}\begin{bmatrix}\begin{array}{r}
-i\\1\\1\\\frac{1}{\sqrt{3}}\\\frac{1}{\sqrt{3}}\\\frac{1}{\sqrt{3}}\end{array}
\end{bmatrix},\quad
\ket{\tilde{\lambda}^{-i}} = \frac{1}{2}\begin{bmatrix}\begin{array}{r}
i\\1\\-1\\\frac{1}{\sqrt{3}}\\\frac{1}{\sqrt{3}}\\\frac{1}{\sqrt{3}}\end{array}
\end{bmatrix},\\
&\ket{\nu_1} = \frac{1}{\sqrt{2}}\begin{bmatrix}\begin{array}{r}
0\\0\\0\\1\\-1\\0\end{array}
\end{bmatrix},\quad
\ket{\nu_2} = \frac{1}{\sqrt{6}}\begin{bmatrix}\begin{array}{r}
0\\0\\0\\1\\1\\-2\end{array}
\end{bmatrix}.
\end{align*}

For comparison, we can show the eigenvectors of the original SQW on the star graph $S_3$, described by the evolution operator $U$ in Eq.~(\ref{eq:start3U}):
\begin{align}\label{eq:spectrastar3}
&\ket{{\lambda}^{+i}} = \frac{1}{2}\begin{bmatrix}\begin{array}{r}
-1\\-1\\i\\1\end{array}
\end{bmatrix},\quad
\ket{{\lambda}^{-i}} = \frac{1}{2}\begin{bmatrix}\begin{array}{r}
1\\-1\\-i\\1\end{array}
\end{bmatrix},\quad
\ket{{\lambda}^{+1}} = \frac{1}{2}\begin{bmatrix}\begin{array}{r}
-i\\1\\1\\1\end{array}
\end{bmatrix},\quad
\ket{{\lambda}^{-i}} = \frac{1}{2}\begin{bmatrix}\begin{array}{r}
i\\1\\-1\\1\end{array}
\end{bmatrix}.
\end{align}
We can see that the construction of the eigenvectors $\ket{\tilde{\lambda}}$ of $\tilde{U}$ follow the rules described in Section~\ref{sec:spectra} and $\tilde{U}$ has two additional $-1$-eigenvectors, $\ket{\nu_1}$ and $\ket{\nu_2}$, because we have 3 tessellations. They can be found directly from Eq.~(\ref{eq:nuj}).

\section{Intersection Reduction}\label{sec:red}
Now consider the opposite case. We have a SQW on a graph which has more than one vertex on some intersection of polygons. In this case, we can show that we can ``reduce" the intersection to one vertex, depending on some conditions. Consider that we have a SQW on a graph $\tilde{G} = (\tilde{V},\tilde{E})$ with $l$ tessellations. Suppose that there are $k$ vertices in the intersection ($I = \{0,1,\dots,k-1\}$) of polygons $\tilde{P_j}$, $0\leq j\leq l-1$.
Let the states induced by the polygons $\tilde{P_j}$ be defined as
\begin{equation}
\ket{\tilde{P}_j} = \sum_{v\in \tilde{V}} \tilde{c_v}^{(j)}\ket{v}.
\end{equation}
In order to be able to reduce the intersection, the amplitudes of the vertices in the intersection in the states induced by polygons should be multiples of each other, that is, $\tilde{c}_v^{(j)} = \beta_{ji}\tilde{c}_v^{(i)} \quad\forall v \in I$.

In this way, we can define
\begin{equation}\label{eq:tildeu}
\ket{\tilde{u}} = \frac{1}{\sqrt{\sum_{v\in I}\left|\tilde{c}_v^{(j)}\right|^2}}\sum_{v\in I}\tilde{c}_v^{(j)}\ket{v},
\end{equation}
for some tessellation $j$. You can choose any tessellation.  
Then, we can rewrite
\begin{equation}
\ket{\tilde{P}_j} = \sum_{v\in \tilde{V}\backslash I}\tilde{c}_v^{(j)}\ket{v}+\tilde{c}_u^{(j)}\ket{\tilde{u}},
\end{equation}
where $\tilde{c}_u^{(j)} = \frac{\tilde{c}_v^{(j)}}{u_v}$.

We can see that the evolution operator $\tilde{U}$ is described exactly as in Section~\ref{sec:exp}. Therefore, its eigenvalues and eigenvectors and the ones from the reduced operator $U$ obey the construction described in Section~\ref{sec:spectra}, with $\tilde{U}$ having additional $k-1$ eigenvectors associated to eigenvalues $\pm 1$ depending on the tessellation number.

\subsection{Example - Reduction to star graph $S_3$}
Suppose we have a SQW on the graph depicted in Fig.~\ref{fig:star2}. In this case, the vertices in the intersection $I = \{3,4,5\}$. Let the unit vectors associated to the polygons in the blue tessellation be
\begin{align}\label{eq:tildealphas}
\ket{\tilde{\alpha}_0} = \frac{1}{\sqrt{2}}\ket{0}-\frac{1}{\sqrt{6}}\ket{3}-\frac{i}{2\sqrt{3}}\ket{4}+\frac{i}{2}\ket{5},\quad \ket{\tilde{\alpha}_1} = \ket{1},\quad \ket{\tilde{\alpha}_2} = \ket{2}.
\end{align}
For the red tessellation we have,
\begin{align}\label{eq:tildebetas}
\ket{\tilde{\beta}_0} = \ket{0},\quad \ket{\tilde{\beta}_1} = \frac{1}{\sqrt{2}}\ket{1}+\frac{i}{\sqrt{3}}\ket{3}-\frac{1}{2\sqrt{3}}\ket{4}+\frac{1}{2}\ket{5},\quad \ket{\tilde{\beta}_2} = \ket{2}.
\end{align}
And for the green tessellation,
\begin{align}\label{eq:tildegammas}
\ket{\tilde{\gamma}_0} = \ket{0},\quad \ket{\tilde{\gamma}_1} = \ket{1},\quad \ket{\tilde{\gamma}_2} = \frac{1}{\sqrt{2}}\ket{2}+\frac{i}{\sqrt{3}}\ket{3}-\frac{1}{2\sqrt{3}}\ket{4}+\frac{1}{2}\ket{5}.
\end{align}
The local operators are
$$
\tilde{U}_{\textrm{blue}} = 2\sum_{i = 0}^2\ket{\tilde{\alpha}_i}\bra{\tilde{\alpha}_i}-I,\quad \tilde{U}_{\textrm{red}} = 2\sum_{i = 0}^2\ket{\tilde{\beta}_i}\bra{\tilde{\beta}_i}-I,\quad 
\tilde{U}_{\textrm{green}} = 2\sum_{i = 0}^2\ket{\tilde{\gamma}_i}\bra{\tilde{\gamma}_i}-I
$$
and the evolution operator $\tilde{U} = \tilde{U}_{\textrm{green}} \tilde{U}_{\textrm{red}}\tilde{U}_{\textrm{blue}}$.

We can see that $\tilde{c}_v^{(i)} = \beta_{ij}\tilde{c}_v^{(j)} \quad\forall v \in I$. For example, $\tilde{c}_3^{(0)} = -i\tilde{c}_3^{(1)}=-i\tilde{c}_3^{(2)}$. Then we can find an expression for $\ket{\tilde{u}}$, from Eq.~(\ref{eq:tildeu}). If we use $j=1$, then
\begin{align*}
\ket{\tilde{u}} = \frac{i}{\sqrt{3}}\ket{3}-\frac{1}{\sqrt{6}}\ket{4}+\frac{i}{\sqrt{2}}\ket{5}.
\end{align*}
We can now rewrite states $\ket{\tilde{\alpha_0}}$, $\ket{\tilde{\beta_1}}$, $\ket{\tilde{\gamma_2}}$ as
$$
\ket{\tilde{\alpha_0}} = \frac{1}{\sqrt{2}}\left(\ket{0}+i\ket{\tilde{u}}\right),\quad \ket{\tilde{\beta_1}} = \frac{1}{\sqrt{2}}\left(\ket{1}+\ket{\tilde{u}}\right),\quad \ket{\tilde{\gamma_2}} = \frac{1}{\sqrt{2}}\left(\ket{2}+\ket{\tilde{u}}\right).
$$
This is a slightly different walk as the one presented in Section~\ref{sec:expexample}. In this case, the evolution operator is
\begin{align*}
\tilde{U} = i\ket{\tilde{\lambda}^{+i}}\bra{\tilde{\lambda}^{+i}}-i\ket{\tilde{\lambda}^{-i}}\bra{\tilde{\lambda}^{-i}}+\ket{\tilde{\lambda}^{+1}}\bra{\tilde{\lambda}^{+1}}-\left(\ket{\tilde{\lambda}^{-1}}\bra{\tilde{\lambda}^{-1}}+\ket{\nu_1}\bra{\nu_1}+\ket{\nu_2}\bra{\nu_2}\right),
\end{align*}
where
\begin{align*}
&\ket{\tilde{\lambda}^{+i}} = \frac{1}{2}\begin{bmatrix}\begin{array}{r}
-1\\-1\\i\\\frac{i}{\sqrt{3}}\\-\frac{1}{\sqrt{6}}\\\frac{1}{\sqrt{2}}\end{array}
\end{bmatrix},\quad
\ket{\tilde{\lambda}^{-i}} = \frac{1}{2}\begin{bmatrix}\begin{array}{r}
1\\-1\\-i\\\frac{i}{\sqrt{3}}\\-\frac{1}{\sqrt{6}}\\\frac{1}{\sqrt{2}}\end{array}
\end{bmatrix},\quad
\ket{\tilde{\lambda}^{+1}} = \frac{1}{2}\begin{bmatrix}\begin{array}{r}
-i\\1\\1\\\frac{i}{\sqrt{3}}\\-\frac{1}{\sqrt{6}}\\\frac{1}{\sqrt{2}}\end{array}
\end{bmatrix},\quad
\ket{\tilde{\lambda}^{-i}} = \frac{1}{2}\begin{bmatrix}\begin{array}{r}
i\\1\\-1\\\frac{i}{\sqrt{3}}\\-\frac{1}{\sqrt{6}}\\\frac{1}{\sqrt{2}}\end{array}
\end{bmatrix},\\
&\ket{\nu_1} = \frac{1}{\sqrt{3}}\begin{bmatrix}\begin{array}{r}
0\\0\\0\\-i\\-\sqrt{2}\\0\end{array}
\end{bmatrix},\quad
\ket{\nu_2} = \frac{1}{\sqrt{6}}\begin{bmatrix}\begin{array}{r}
0\\0\\0\\i\sqrt{2}\\-1\\-\sqrt{3}\end{array}
\end{bmatrix}.
\end{align*}

Now we reduce the intersection to one vertex. Let us label the new vertex as $3$. Our new graph $G=(V,E)$ has vertex set $V=\{0,1,2,3\}$ and we obtain exactly the SQW on Fig.~\ref{fig:star}. The unit vectors associated to the polygons will be the ones described for the graph $\tilde{G}$ by substituting $\ket{\tilde{u}}$ by $\ket{3}$. This gives us exactly the states in Eqs.~(\ref{eq:sqwstarblue})-(\ref{eq:sqwstargreen}). The eigenvectors for this SQW is shown in Eq.~(\ref{eq:spectrastar3}). From this example we can see how the same SQW $U$ can generate different $\tilde{U}$'s depending on how we define $\ket{\tilde{u}}$.

\section{Results}\label{sec:results}

We can combine our results in the following theorems. Let $U$ and $\tilde{U}$ be evolution operators from SQWs associated to graphs $G=(V,E)$ and $\tilde{G} = (\tilde{V},\tilde{E})$, respectively, obtained from an intersection expansion or reduction process as described in Sections~\ref{sec:exp} and \ref{sec:red}. 

\begin{theorem}\label{teo1}
 Let $\ket{\psi} \in \mathcal{H}^{|V|}$ be a generic state and $\ket{\tilde{\psi}} \in \mathcal{H}^{|\tilde{V}|}$ be defined as in Eq.~(\ref{eq:tildepsi}). 
The action of the local operators on these vectors, that is, $U_j\ket{\psi}$ and $\tilde{U}\ket{\psi}$, preserves the amplitudes of the vertices that don't belong to the reduced or expanded intersection and preserves the probability of obtaining a vertex in the intersection after measurement in the computational basis, that is, $\braket{v|\tilde{U}_j|\tilde{\psi}} = \braket{v|{U}_j|{\psi}}$, for all $v\in V\backslash \{u\}$ and
$\braket{\tilde{u}|\tilde{U}_j|\tilde{\psi}} = \braket{{u}|{U}_j|{\psi}}$.
\end{theorem}

\begin{theorem}\label{theorem}
$\tilde{U}$ has $k-1$ eigenvectors $\ket{\nu_j}, 1\leq j\leq k-1$, given by Eq.~(\ref{eq:nuj}), where $k$ is the number of vertices in the intersection of some polygons. They are associated with eigenvalue $-1$ if the number of tessellations is odd. Otherwise, they are associated with eigenvalue $+1$. The remaining eigenvalues and eigenvectors of $U$ and $\tilde{U}$ are as follows. $\ket{\lambda}$ (see Eq.~(\ref{eq:lambda})) is a $\lambda$-eigenvector of $U$ and $\ket{\tilde{\lambda}}$ (given by Eq.~(\ref{eq:tildelambda})) is also a $\lambda$-eigenvector of $\tilde{U}$. 
\end{theorem}

Notice that the intersection reduction or expansion process can be applied on multiple intersections one at a time, as we can see in the following example.

\subsection{Example - Search on the two dimensional lattice}\label{sec:examplesearch}

\begin{figure}[!htb]
  \centering
  \begin{subfigure}[t]{0.4\textwidth}
  \centering
\includegraphics[scale=1.2]{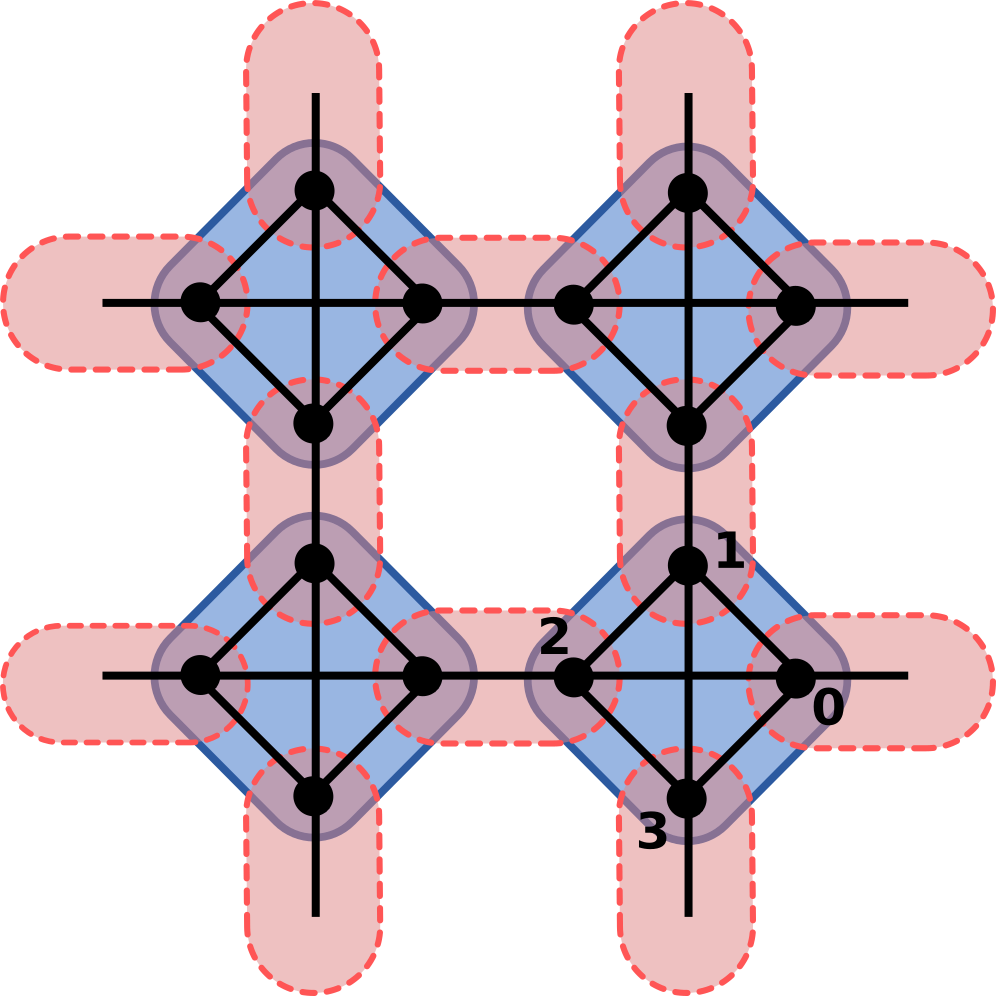}  
     \caption{The SQW equivalent of the flip-flop coined quantum walk on the two dimensional lattice.}
     \label{fig:4-clique-grid}
 \end{subfigure}
 \hspace{5mm}
 \begin{subfigure}[t]{0.4\textwidth}
 \centering
 \includegraphics[scale=1.2]{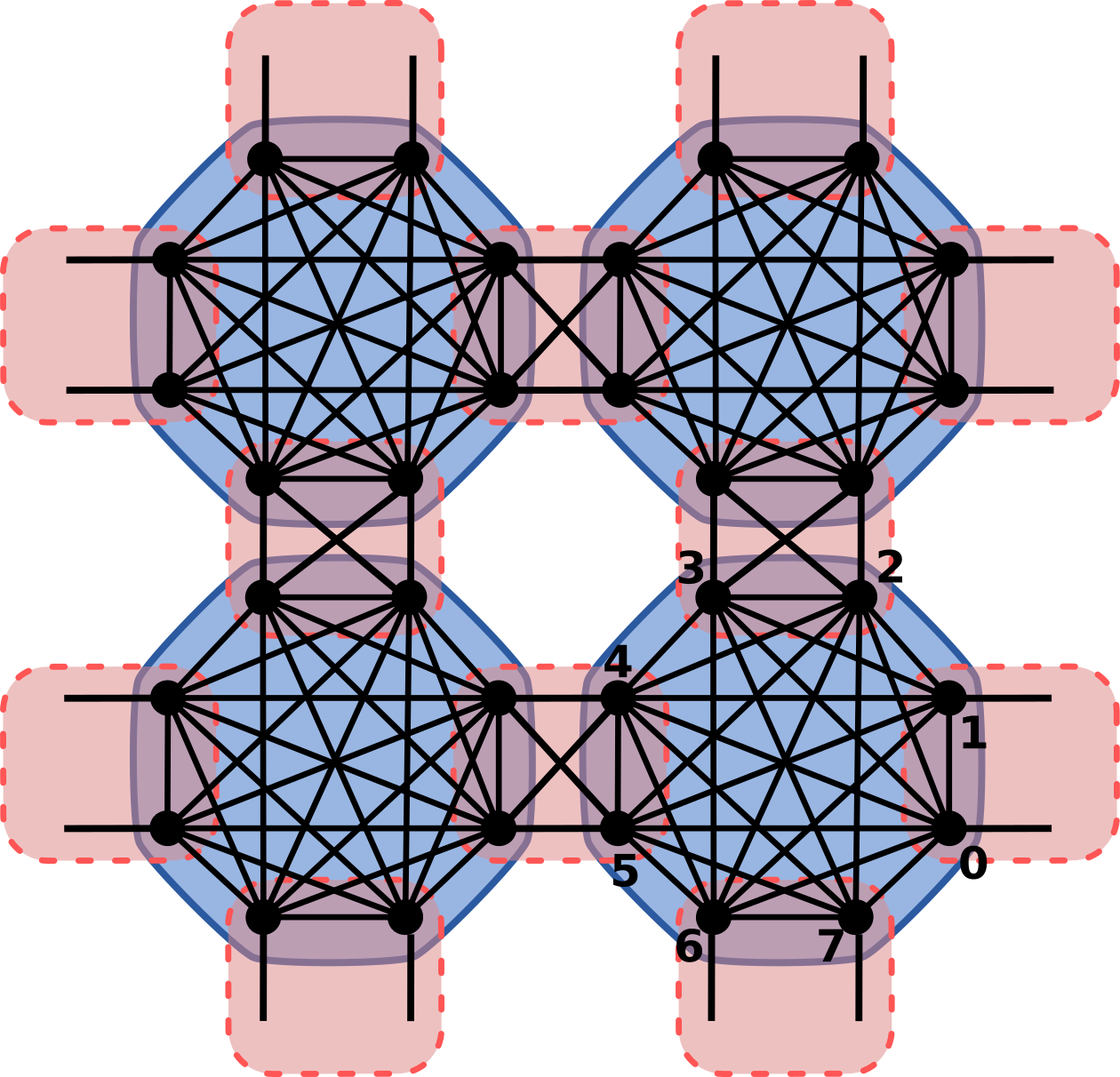}
  \caption{Substituting each vertex in \ref{fig:4-clique-grid} by two vertices.}
  \label{fig:8-clique-grid}
 \end{subfigure}\\
 \begin{subfigure}[t]{0.5\textwidth}
 \centering
 \includegraphics[scale=1.2]{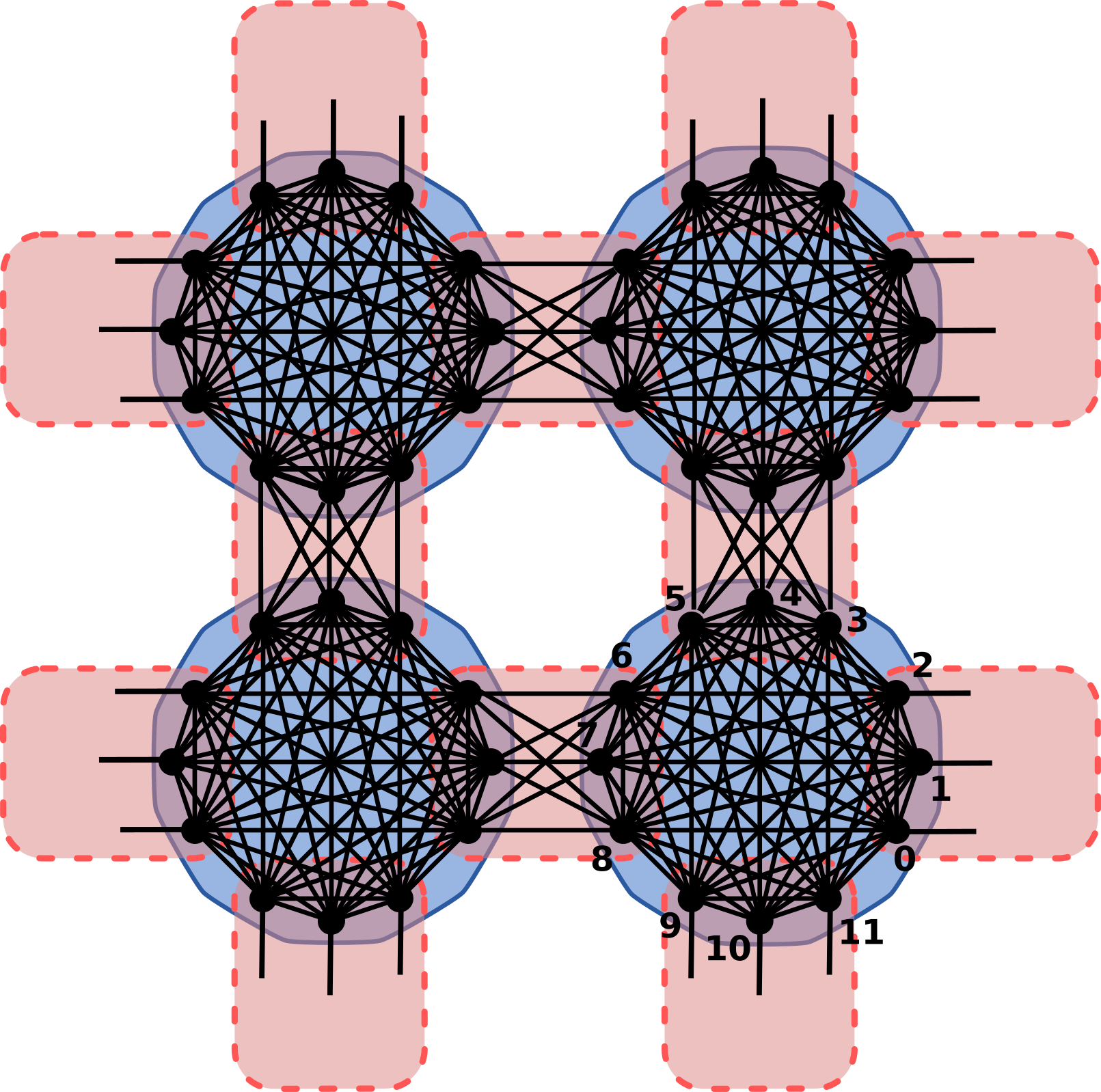}
  \caption{Substituting each vertex in \ref{fig:4-clique-grid} by tree vertices.}
  \label{fig:12-clique-grid}
 \end{subfigure}
 \caption{SQWs on the equivalent of the flip-flop coined QW on the two dimensional lattice and on some intersection expansions.}
 \label{fig:sqw}
\end{figure}
 A standard flip-flop coined QW with the four-dimensional Grover coin on the
two dimensional lattice is equivalent to the SQW on the graph and tessellation partition depicted in Fig.~\ref{fig:4-clique-grid} \cite{Portugal:2016}. This graph is obtained by substituting each vertex of a $n\times n$ two dimensional grid by a $4$-clique.

Portugal~\cite{Portugal:2016a} numerically analyzed the search problem in the graph depicted in Fig.~\ref{fig:8-clique-grid}. This graph is obtained by substituting each vertex in graph~\ref{fig:4-clique-grid} by two vertices. It consists of $n^2$ 8-cliques linked by $2n^2$ 4-cliques with a torus-like topology, as in graph~\ref{fig:4-clique-grid}. 
This graph is in the class of graphs that are not line graphs. See Ref.~\cite{Portugal:2016a} for more details. What is important about this class is that the 2-tessellable SQWs on graphs in this class have one or more
edges in the intersection of the tessellations and SQWs on graphs in this class can be included neither in Szegedy's model nor in the flip-flop coined model.

The graphs in Fig.~\ref{fig:8-clique-grid} and \ref{fig:12-clique-grid} are obtained by substituting each vertex in the graph of Fig.~\ref{fig:4-clique-grid} by 2 vertices and by 3 vertices, respectively. We will prove, using our results, that searching for a clique on the blue polygons in any of these graphs is equivalent to searching for a vertex in the two dimensional lattice.

Suppose we increase each vertex in graph~\ref{fig:4-clique-grid} by $q$ vertices, where $q\geq 1$, then the graph will consist of $n^2$ $4q$-cliques linked by $2n^2$ $2q$-cliques with a torus-like topology. Graph in Fig.~\ref{fig:12-clique-grid} is obtained for $q=3$.
We can define a SQW on these graphs. The Hilbert space associated to the graph has dimension $4qn^2$.
The vectors associated with the blue polygons are
\begin{equation}
\ket{\tilde{\alpha}_{xy}} = \frac{1}{2\sqrt{q}}\sum_{k=0}^{4q-1}\ket{x,y,k},
\end{equation}
and the vectors associated with the red polygons are
\begin{equation}
\begin{split}
\ket{\tilde{\beta}_{xy}^{(0)}} &= \frac{1}{\sqrt{2q}}\sum_{k=0}^{q-1}\left(\ket{x,y}\ket{k}+\ket{x+1,y}\ket{2q+k}\right),\\
\ket{\tilde{\beta}_{xy}^{(1)}} &=\frac{1}{\sqrt{2q}}\sum_{k=0}^{q-1}\left(\ket{x,y}\ket{q+k}+\ket{x,y+1}\ket{3q+k}\right),
\end{split}
\end{equation}
for $0\leq x,y \leq n-1$ and the arithmetic with the labels of $\ket{x,y}$ is performed modulo $n$. 

The evolution operator is $\tilde{U} = \tilde{U}_1\tilde{U}_0$, where 
\begin{equation}
\tilde{U}_0 = 2\sum_{x,y = 0}^{n-1}\ket{\tilde{\alpha}_{xy}}\bra{\tilde{\alpha}_{xy}} - I,
\end{equation}
and
\begin{equation}
\tilde{U}_1 = 2\sum_{x,y = 0}^{n-1}\ket{\tilde{\beta}_{xy}^{(0)}}\bra{\tilde{\beta}_{xy}^{(0)}}+\ket{\tilde{\beta}_{xy}^{(1)}}\bra{\tilde{\beta}_{xy}^{(1)}} - I,
\end{equation}

There are $4n^2$ intersections of polygons with $q$ vertices. 

\subsubsection{Intersection reduction}
\begin{figure}[!htb]
  \centering
\includegraphics[scale=1.2]{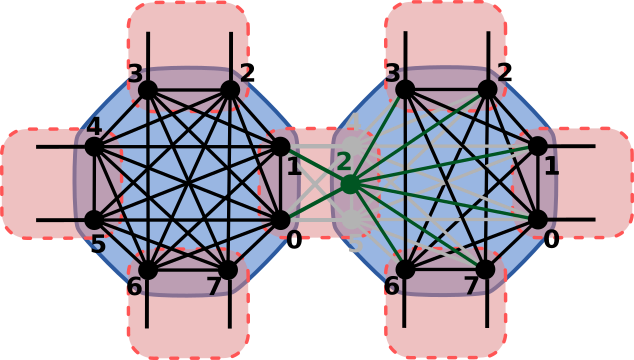}  
     \caption{Example of reducing the intersection of two polygons (which contains vertices 4 and 5) to one vertex, in the case $q=2$.}
     \label{fig:redgrid}
\end{figure}

By Theorem~\ref{theorem}, we can reduce each intersection to one vertex. For doing that we can easily verify that the amplitudes in the states $\ket{\tilde{\alpha}_{xy}}$, $\ket{\tilde{\beta}_{xy}^{(0)}}$, $\ket{\tilde{\beta}_{xy}^{(1)}}$ satisfy the necessary conditions (see Section~\ref{sec:red}). 
We can find the states associated to each one of the intersections, which are
$$
\ket{\tilde{u}_{x,y}^{(m)}} = \frac{1}{\sqrt{q}}\sum_{k=0}^{q-1}\ket{x,y,mq+k}
$$
where $0\leq m\leq 3$. Then, we can rewrite
\begin{equation}
\ket{\tilde{\alpha}_{xy}} = \frac{1}{2}\sum_{k=0}^{4}\ket{\tilde{u}_{xy}^{(k)}}
\end{equation}
and \begin{equation}
\begin{split}
\ket{\tilde{\beta}_{xy}^{(0)}} &= \frac{1}{\sqrt{2}}\ket{\tilde{u}_{x,y}^{(0)}}+\ket{\tilde{u}_{x+1,y}^{(2)}}\\
\ket{\tilde{\beta}_{xy}^{(1)}} &=\frac{1}{\sqrt{2}}\ket{\tilde{u}_{x,y}^{(1)}}+\ket{\tilde{u}_{x,y+1}^{(3)}}.
\end{split}
\end{equation}
Now we can substitute the vertices in each intersection by one vertex. Let us name vertices $\{ mq+k | k=0,\dots,q-1 \}$ in each $4q$-clique as vertex $m$ in the reduced graph. 
Fig.~\ref{fig:redgrid} shows how this process is done for one intersection when $q=2$.
This process is applied to all intersections and we obtain exactly the SQW on the graph given by Fig.~\ref{fig:4-clique-grid}, that is,
\begin{equation}
\ket{\alpha_{xy}} = \frac{1}{2}\sum_{k=0}^{4}\ket{x,y,k}
\end{equation}
and the vectors associated with the red polygons are
\begin{equation}
\begin{split}
\ket{\beta_{xy}^{(0)}} &= \frac{1}{\sqrt{2}}\ket{x,y,0}+\ket{x+1,y,2}\\
\ket{\beta_{xy}^{(1)}} &=\frac{1}{\sqrt{2}}\ket{x,y,1}+\ket{x,y+1,3},
\end{split}
\end{equation}
The new evolution operator $U$ is the operator $\tilde{U}$ for $q=1$.

\subsubsection{Spatial Search}

The implementation of spatial search on SQWs can be done by using partial tessellations. This is simply done by removing polygons from the tessellation and the vertices in the missing polygons will be the marked ones.  
\begin{figure}[!htb]
  \centering
\includegraphics[scale=1.2]{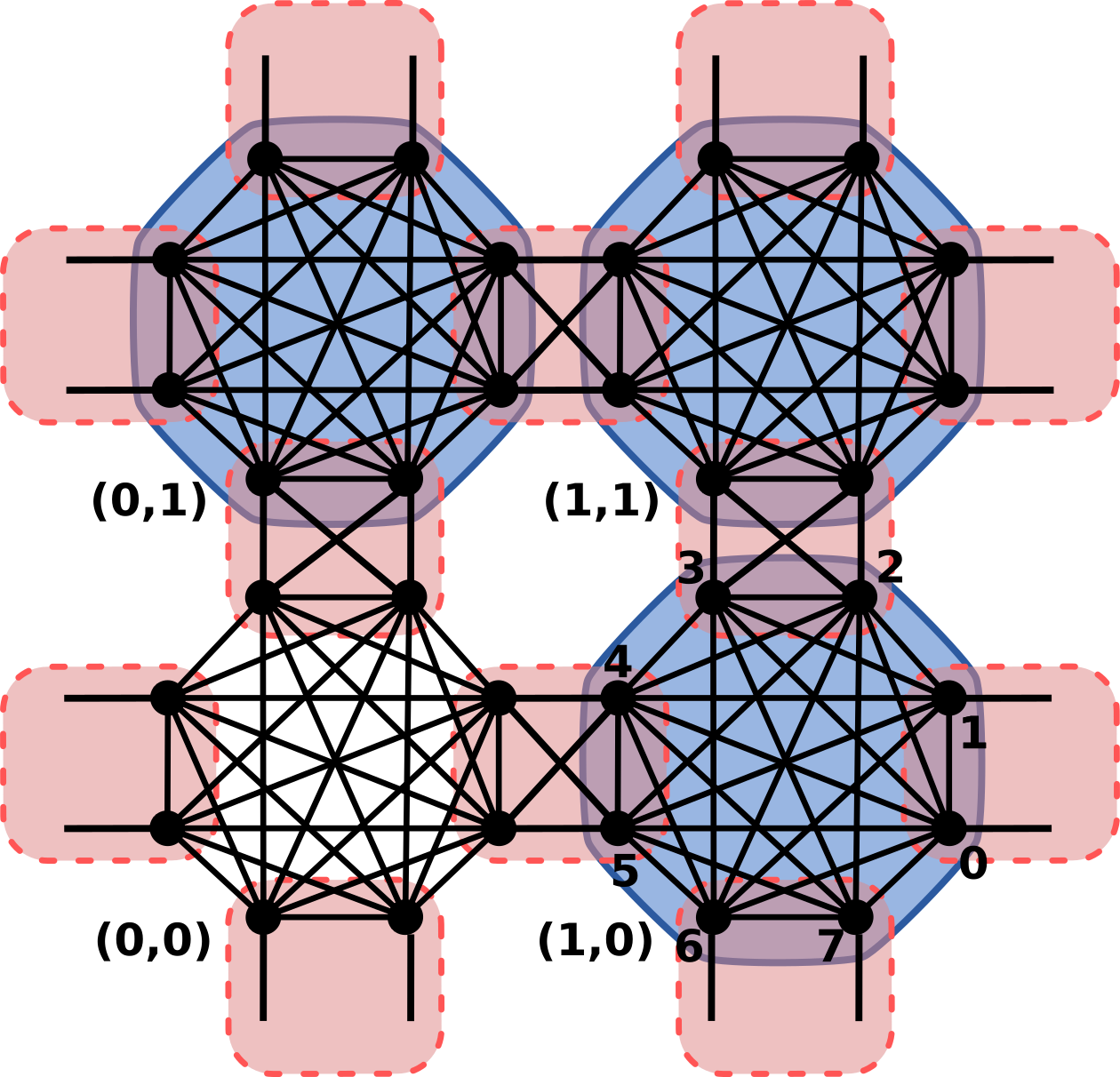}  
     \caption{Search for the 8-clique in position $(0,0)$. The blue polygon containing the clique is removed and we have a SQW with a partial tessellation.}
     \label{fig:sqwsearch}
\end{figure}
Without loss of generality, let us assume the $4q$-clique in position $(0,0)$ is the marked one. Then, the search operator $\tilde{U}'$ is obtained by removing the polygon which induces the state $\ket{\tilde{\alpha}_{00}}$, as we can see in Fig.~\ref{fig:sqwsearch}. The evolution operator is $\tilde{U}' = \tilde{U}_1 \tilde{U}'_0$, where
$$
\tilde{U}'_0 = \left(2\sum_{\substack{x,y = 0\\(x,y)\neq (0,0)}}^{n-1}\ket{{\tilde{\alpha}}_{xy}}\bra{{\tilde{\alpha}}_{xy}} - I\right).
$$
The initial state is the uniform superposition of all vertices of the graph,
$$
\ket{\psi(0)} = \frac{1}{|\tilde{V}|}\sum_{v\in \tilde{V}}\ket{v}.
$$

It is possible to show that the SQW $\tilde{U}$ has the same complexity as the SQW $U$ for searching a marked clique in the blue polygons.
An intuitive proof of that fact is by showing that $\tilde{U}'$ acts the same as $U'$, preserving the same probability in both cliques during time.
In Section~\ref{sec:exp}, we have showed that for some $\tilde{U}$ and $U$. But now we have to consider the search algorithm.
The difference between $U'_0/\tilde{U}'_0$ and $U_0/\tilde{U}_0$ is how they act on the missing polygon. $U'_0$ and $\tilde{U }'_0$ will simply flip the sign of the amplitudes of all vertices in the marked clique, that is,
\begin{align*}
&U'_0\left(\frac{1}{2}\sum_{k=0}^{3}\ket{0,0,k}\right) = -\frac{1}{2}\sum_{k=0}^{3}\ket{0,0,k},\\
&\tilde{U}'_0\left(\frac{1}{2\sqrt{q}}\sum_{k=0}^{4q-1}\ket{0,0,k}\right) = -\frac{1}{2\sqrt{q}}\sum_{k=0}^{4q-1}\ket{0,0,k} = -\frac{1}{2}\sum_{m=0}^3\ket{\tilde{u}_{0,0}^{(m)}}.
\end{align*}

The initial state can be written in terms of $\ket{\tilde{u}_{x,y}^{(m)}}$, then the probability of obtaining a marked vertex (of $U'$ or $\tilde{U}'$) will be the same throughout the evolution. From Theorem~\ref{teo1}, the other intersections which does not contain marked vertices will preserve the same probability.
Ref.~\cite{Portugal:2016} showed that the SQW with a missing blue polygon on graph~\ref{fig:4-clique-grid} is equivalent to
the non-regular flip-flop coined QW on the two-dimensional lattice with coin ($−I$) on the
marked vertex and the Grover coin on the non-marked vertices. 
Therefore $U'$ and $\tilde{U}'$ have the same probability of success as for searching a marked vertex on the two dimensional lattice. The number of steps of the algorithm is $O(\sqrt{N\log N})$ and the success probability is $O(1/\log N)$, where $N=n^2$. The total cost of the algorithm applying the amplitude amplification method is $O(\sqrt{N}\log N)$.  

Another way to prove is to use the abstract search algorithm scheme~\cite{Ambainis:2005} and find the spectral decomposition of $\tilde{U}$, which can be obtained from $U$ by Theorem~\ref{theorem}.

\section{Conclusions}\label{sec:conc}


We have analyzed the role that the size of the polygon intersection plays on the dynamics of SQWs on graphs.
We have introduced an intersection reduction and expansion processes, which decreases or increases the number of vertices in some intersection of polygons. We have showed how the SQW on the reduced or expanded graph behaves in relation to the SQW on the original graph. 
 We also have described how the eigenvectors and eigenvalues of the reduced or expanded SQW relates to the eigenvectors and eigenvalues of the original SQW. We can conclude that both SQWs are equivalent if the vertices in the intersection are not treated individually.

From the example in Sec.~\ref{sec:examplesearch},  we can see that it is possible to find SQWs that are not in Szegedy's model and which are equivalent to an instance of Szegedy's model after an intersection reduction process.
Moreover, it is also possible to use an intersection reduction or expansion process on SQW-based search algorithms, depending on the location of the marked vertices.

Since an intersection of polygons with more than one vertex is responsible for $\pm 1$-eigenvectors, when the conditions for the reduction process are satisfied, it would be interesting to study how localization plays a role on SQWs with more than one vertex in the intersection of polygons.

\section*{Acknowledgements}
The author thanks Renato Portugal for useful discussions and comments, and Bruno Chagas for sharing some calculations on SQWs.
This work was suported by ERDF project number 1.1.1.2/VIAA/1/16/002.

\bibliographystyle{unsrt}
\bibliography{refs}

\begin{thebibliography}{10}

\bibitem{Portugal:2015}
R.~Portugal, R.~A.~M. Santos, T.~D. Fernandes, and D.~N. Gon{\c{c}}alves.
\newblock The staggered quantum walk model.
\newblock {\em Quantum Information Processing}, 15(1):85--101, 2015.

\bibitem{Szegedy:2004}
M.~Szegedy.
\newblock Quantum speed-up of markov chain based algorithms.
\newblock In {\em Proceedings of the 45th Symposium on Foundations of Computer
  Science}, pages 32--41, 2004.

\bibitem{Portugal:2016}
Renato Portugal.
\newblock Establishing the equivalence between \uppercase{S}zegedy's and coined
  quantum walks using the staggered model.
\newblock {\em Quantum Information Processing}, pages 1--23, 2016.

\bibitem{Aharonov:1993}
Y.~Aharonov, L.~Davidovich, and N.~Zagury.
\newblock Quantum random walks.
\newblock {\em Physical Review A}, 48(2):1687--1690, 1993.

\bibitem{Aharonov:2001}
Dorit Aharonov, Andris Ambainis, Julia Kempe, and Umesh Vazirani.
\newblock Quantum walks on graphs.
\newblock In {\em Proceedings of the Thirty-third Annual ACM Symposium on
  Theory of Computing}, STOC '01, pages 50--59, New York, NY, USA, 2001. ACM.

\bibitem{Ambainis:2005}
A.~Ambainis, J.~Kempe, and A.~Rivosh.
\newblock Coins make quantum walks faster.
\newblock In {\em Proceedings of the 16th ACM-SIAM Symposium on Discrete
  Algorithms}, pages 1099--1108, 2005.

\bibitem{Coutinho:2018}
Gabriel Coutinho and Renato Portugal.
\newblock Discretization of continuous-time quantum walks via the staggered
  model with hamiltonians.
\newblock {\em Natural Computing}, Jun 2018.

\bibitem{Farhi:1998}
E.~Farhi and S.~Gutmann.
\newblock Quantum computation and decision trees.
\newblock {\em Physical Review A}, 58:915--928, 1998.

\bibitem{Portugal:2017}
R.~Portugal, M.~C. de~Oliveira, and J.~K. Moqadam.
\newblock Staggered quantum walks with hamiltonians.
\newblock {\em Phys. Rev. A}, 95:012328, Jan 2017.

\bibitem{Jalil:2017}
J.~Khatibi~Moqadam, M.~C. de~Oliveira, and R.~Portugal.
\newblock Staggered quantum walks with superconducting microwave resonators.
\newblock {\em Phys. Rev. B}, 95:144506, Apr 2017.

\bibitem{Portugal:2017a}
R.~Portugal and T.~D. Fernandes.
\newblock Quantum search on the two-dimensional lattice using the staggered
  model with hamiltonians.
\newblock {\em Phys. Rev. A}, 95:042341, Apr 2017.

\bibitem{Chagas:2018}
Bruno~A. Chagas, Renato Portugal, Stefan Boettcher, and Etsuo Segawa.
\newblock Staggered quantum walk on hexagonal lattices.
\newblock \url{arXiv/quant-ph:1806.10249}, 2018.

\bibitem{Portugal:2018}
Renato Portugal.
\newblock Element distinctness revisited.
\newblock {\em Quantum Information Processing}, 17(7):163, May 2018.

\bibitem{Ambainis:2004}
A.~Ambainis.
\newblock Quantum walk algorithm for element distinctness.
\newblock In {\em Proceedings of the 45th Annual IEEE Symposium on Foundations
  of Computer Science}, 2004.

\bibitem{Abreu:2018}
Alexandre Abreu, Lu{\'i}s Cunha, Tharso Fernandes, Celina de~Figueiredo, Luis
  Kowada, Franklin Marquezino, Daniel Posner, and Renato Portugal.
\newblock The graph tessellation cover number: Extremal bounds, efficient
  algorithms and hardness.
\newblock In Michael~A. Bender, Mart{\'i}n Farach-Colton, and Miguel~A.
  Mosteiro, editors, {\em LATIN 2018: Theoretical Informatics}, pages 1--13,
  Cham, 2018. Springer International Publishing.

\bibitem{Portugal:2016a}
Renato Portugal.
\newblock Staggered quantum walks on graphs.
\newblock {\em Phys. Rev. A}, 93:062335, Jun 2016.

\end{thebibliography}

\end{document}